%% file: main.tex
\documentclass[lettersize,journal]{IEEEtran}
\usepackage{amsmath,amsfonts}
\usepackage{algorithmic}
\usepackage{algorithm}
\usepackage{array}
\usepackage[caption=false,font=normalsize,labelfont=sf,textfont=sf]{subfig}
\usepackage{textcomp}
\usepackage{stfloats}
\usepackage{url}
\usepackage{verbatim}
\usepackage{graphicx}
\usepackage{cite}
\usepackage{bm}
\usepackage{placeins} 
\usepackage{color}
\usepackage{placeins}
\usepackage[compact]{titlesec}
\titlespacing*{\section}{0pt}{*0.7}{*0.5}
\titlespacing*{\subsection}{0pt}{*0.5}{*0.3}

\begin{document}

\title{A Case for Kolmogorov-Arnold Networks in Prefetching: Towards Low-Latency, Generalizable ML-Based Prefetchers}



\author{Dhruv Kulkarni\IEEEauthorrefmark{1}, Bharat Bhammar\IEEEauthorrefmark{1}, Henil Thaker\IEEEauthorrefmark{1}, Pranav Dhobi\IEEEauthorrefmark{1}, \\ 
R.P. Gohil\IEEEauthorrefmark{1}, Sai Manoj Pudukotai Dinkarrao, \textit{Senior Member, IEEE}\IEEEauthorrefmark{2}%
\thanks{\IEEEauthorrefmark{1}The authors are affiliated with the CSE Department at the Sardar Vallabhbhai National Institute of Technology, Surat, Gujarat, India. E-mail: \{u21cs036, u21cs065,  u21cs039, u21cs050, rpg\}@coed.svnit.ac.in. }
\thanks{\IEEEauthorrefmark{2}Dr. Sai Manoj Pudukotai Dinkarrao is affiliated with Department of Electrical and Computer Engineering, George Mason University, Fairfax, VA 22030, USA. E-mail: spudukot@gmu.edu.}
}


\maketitle
\setlength{\abovedisplayskip}{3pt}
\setlength{\belowdisplayskip}{3pt}
\setlength{\parskip}{0pt}
\setlength{\textfloatsep}{6pt}
\setlength{\floatsep}{6pt}
\setlength{\intextsep}{6pt}
\setlength{\abovecaptionskip}{3pt}
\setlength{\belowcaptionskip}{2pt}

\begin{abstract}

The memory wall problem arises due to the disparity between fast processors and slower memory, causing significant delays in data access, even more so on edge devices. Data prefetching is a key strategy to address this, with traditional methods evolving to incorporate Machine Learning (ML) for improved accuracy. Modern prefetchers must balance high accuracy with low latency to further practicality. We explore the applicability of utilizing Kolmogorov-Arnold Networks (KAN) with learnable activation functions,a prefetcher we implemented called KANBoost, to further this aim. KANs are a novel, state-of-the-art model that work on breaking down continuous, bounded multi-variate functions into functions of their constituent variables, and use these constitutent functions as activations on each individual neuron. KANBoost predicts the next memory access by modeling deltas between consecutive addresses, offering a balance of accuracy and efficiency to mitigate the memory wall problem with minimal overhead, instead of relying on address-correlation prefetching. Initial results indicate that KAN-based prefetching reduces inference latency (18× lower than state-of-the-art ML prefetchers) while achieving moderate IPC improvements (2.5\% over no-prefetching). While KANs still face challenges in capturing long-term dependencies, we propose that future research should explore hybrid models that combine KAN efficiency with stronger sequence modeling techniques, paving the way for practical ML-based prefetching in edge devices and beyond.

\end{abstract}

\section{Introduction}
\noindent
Advancements in the computing systems and the need to execute data-intensive applications such as machine learning, and high-performance computing 
demand rapid and efficient access to large volumes of data at disposal. However, due to inherent traits of the DRAM and the interconnections, the 
gap between memory access latency and processor speed has become a critical bottleneck. To address this, multiple techniques have been proposed 
in the literature \cite{hennessy2011computer}. Among these techniques, prefetching \cite{sameh1990compiler} plays a pivotal role in mitigating this issue by predicting and fetching data before it is requested, thereby reducing memory access latency and improving overall system performance.

By leveraging advanced prefetching strategies—ranging from hardware-based approaches to machine learning-driven models—modern architectures can enhance data locality, minimize cache misses, and optimize resource utilization, ultimately accelerating computation and improving efficiency in data-driven workloads. 
Multiple prefetching strategies for designing a prefetcher have been proposed in the literature to enhance the performance and minimize the overheads. 

The best offset \cite{Best-offset} prefetcher introduces a sophisticated mechanism that utilizes a priority queue to analyze and interpret different offsets—the magnitude of differences between successive memory accesses. By identifying patterns within these offsets, it predicts the most likely subsequent memory block to be accessed.



In addition to heuristics-based prefetchers, machine learning-based prefetchers are as well introduced in the recent times. 
The Drishyam prefetcher \cite{Drishyam} employs an innovative approach by transforming historical memory block access patterns into image-like representations. Leveraging computer vision models, it classifies these representations to predict the subsequent memory block. Another notable advancement, TransforMAP \cite{zhang2021transformap}, utilizes the Transformer model, renowned for its success in natural language processing and sequence prediction tasks, to forecast the next memory block access. By restricting its predictions to the same memory page, TransforMAP ensures both accuracy and computational efficiency. 

The state-of-the-art prefetcher designs inherits some of the challenges in the existing works and ML techniques: 
1) The inability to understand complex patterns governing memory accesses 2) Lack of generalization ability of ML models underlying these prefetchers. 3) High inference time, making ML-based prefetchers impractical. To address the high inference time/prediction time, we pursue a delta/stride based prefetching strategy, which involves predicting the delta or absolute memory difference between the present and next address, making the predictors less complex and faster. 
To address the issues of understanding complex patterns, we explore a state-of-the-art model, very recently proposed in \cite{liu2024kankolmogorovarnoldnetworks}

In this work, we explore an offset-prefetching strategy, which predicts the next offset based on the history of block accesses, using Kolmogorov-Arnold Networks (KANs) 
\cite{liu2024kankolmogorovarnoldnetworks}, a state-of-the-art learning model, based on the Kolmogorov-Arnold representation, which demonstrates that any bounded, continuous multivariate function can be represented as the summation of the functions of its constituent variables. 
Our preliminary experimental analysis showcases a $18\times$ improvement in prefetching time compared to state-of-the art prefetchers, such as \cite{Voyager}, with up to 2.5\% improvement in IPC on some of the GAP and SPEC benchmarks compared to a no-prefetching scenario.
Our novel contributions can be outlined in a two-fold manner: 
\begin{itemize}
    \item To our knowledge, this is the first work to explore Kolmogorov-arnold Networks and employ learnable activation functions for prefetching in computer architecture. 
    \item This work also introduces a simplified prefetching pipeline. The prefetcher prediction is made lightweight thereby 
    enables predicting the relative address (delta) within the same page, as opposed to generality, in order to make the prefetching feasible.
\end{itemize}




\section{Proposed KANBoost}

\begin{figure*}[htbp]
    \centering
    \includegraphics[width=6 in]{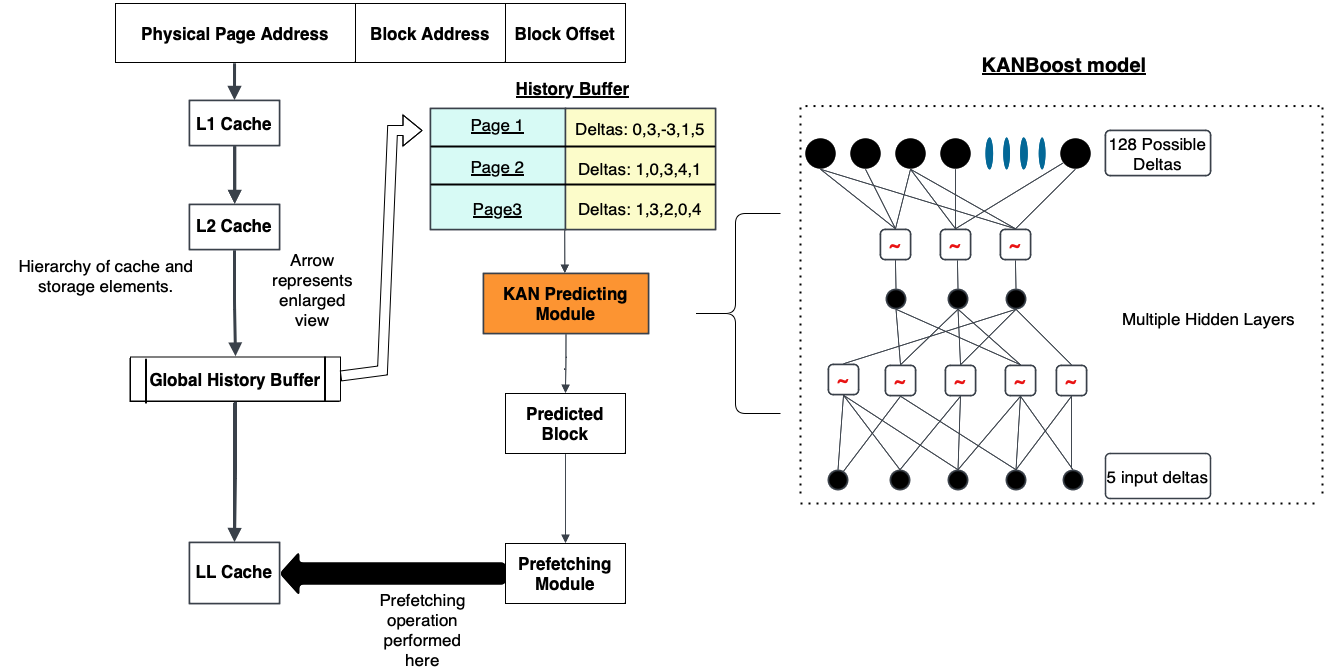} 
    \caption{Prefetching architecture}
    \label{fig:Methodology} 
\end{figure*}

\noindent
In this Section, we discuss the details of the KAN-based prefetcher design. 
The architecture of the proposed prefetcher is illustrated in Figure \ref{fig:Methodology}. 
In contrast to correlation-based prefetchers, this work considers delta-based prefetcher due to its reduced latency in predicting the next address access. 
Figure \ref{fig:Methodology} represents the prefetching architecture. We predict the future memory accesses for the last-level cache (LLC), down the cache hierarchy as demonstrated in Figure \ref{fig:Methodology}. Prefetching is not performed on L1 Cache and L2 Cache.
We collect previous accesses under the same page in the global history buffer and use the KAN predicting module (made up of a KAN model) predict the next block access under the same page.
After concatenating the block address with the page address to form the complete address, the prefetching module prefetches the address to the LLC.


To reduce the sample space for offset prediction, we restrict the prefetcher to predicting offsets within a page for different block accesses. This reduces inference time and additional complexity, in finding block accesses within other pages.

\subsection{Kolmogorov-Arnold Networks}
\noindent
\textit{KAN} \cite{liu2024kankolmogorovarnoldnetworks} is a type of neural network, based on the Kolmogorov-Arnold theorem, which states that any multivariable function can be represented as a finite composition of single-variable functions. This theorem allows \textit{KAN} to decompose complex relationships into simpler sub-problems, thus enhancing accuracy and efficiency in predictions.

The mathematical expression of \textit{KAN} \cite{liu2024kankolmogorovarnoldnetworks} is given by:
\begin{equation}
f(x_1, x_2, \ldots, x_n) = \sum_{i=1}^{2n+1} \phi_i \left( \sum_{j=1}^{n} \psi_{ij}(x_j) \right)
\label{eq:1}
\end{equation} 
\( f(x_1, x_2, \ldots, x_n) \) is the target function that maps the input variables \( x_1, x_2, \dots, x_n \) to the output. \( n \) represents the number of input variables. \( \phi_i(\cdot) \) are learnable activation functions applied to the intermediate summations. \( \psi_{ij}(x_j) \) represents transformation functions applied to each input variable \( x_j \), decomposing the multivariate function.

In the context of memory prefetching, \textit{KAN}'s ability to model intricate relationships between memory accesses makes it a promising candidate for addressing the challenges posed by the memory wall. By capturing the intricacies of memory behavior, \textit{KAN}-based prefetchers like KANBoost can dynamically adjust to changing workloads and optimize memory access in ways that traditional methods may not achieve.

\subsection{Prefetching Pipeline} 
\noindent
 For a computation to process, the data will be prefetched to the LLC, as shown in Figure \ref{fig:Methodology}.
 Consider a simple example :
Let past memory accesses be 1:00, 1:00, 1:03, 1:00, 1:01, and 1:06 (logical page:block representation). From these addresses, deltas, representing the differences between consecutive memory accesses, are calculated (e.g., $\Delta_1 = 0$, $\Delta_2 = 3$, $\Delta_3 = -3$ -- as shown in Figure \ref{fig:Methodology}) (Stage 2 of Algorithm \ref{alg:kanboost}). These deltas are subsequently stored in an array containing the previous $K$ deltas, where $K$ is a configurable parameter. For this example, $K = 5$. This array serves as the input to the KANBoost model, which analyzes the sequence to predict future deltas based on learned patterns (Stages 4 and 5 of \ref{alg:kanboost}). 

Using the predicted deltas, KANBoost generates prefetch addresses. For instance, if the model predicts the next delta as $\Delta_6 = 5$, it computes the subsequent memory address as 1:0B (i.e., 1:06 + $\Delta_6$) and prefetches the corresponding memory block. This proactive approach ensures the data is loaded into memory prior to processor requests, thereby significantly reducing access latency.

The KANBoost prefetching mechanism is structured around three key components:  
\begin{itemize}
    \item \textbf{Delta Computation}: Calculating the differences between consecutive memory addresses to derive a sequence of deltas, as showin in Stage 2 of Algorithm 1.
    \item \textbf{Training dataset}: After preparing the deltas, we prepare a pipeline so as to prepare the training dataset for our KANBoost model, as shown in Stage 3 of Algorithm \ref{alg:kanboost}.
    \item \textbf{KANBoost Model}: The dataset is now utilized for training the ``KANBoost": KAN based predicting module, as demonstrated in Stage 4 of \ref{alg:kanboost}. 
    \item \textbf{Prefetching}: Utilizing predicted deltas to prefetch anticipated memory blocks, thereby minimizing latency by aligning memory access with processor demands.
\end{itemize}

KANBoost’s architecture is designed to capture intricate patterns in memory access behavior, ensuring high adaptability and efficiency. By prefetching memory blocks in advance, the system effectively reduces cache misses and memory latency, resulting in a significant improvement in overall system performance.

The summary of the methodology can be found in Algorithm \ref{alg:kanboost}. 
Preliminary results show that KANBoost is slow in capturing irregular memory accesses. We explain the results in the following sections.

 \begin{algorithm}
 \small
\caption{KANBoost Algorithm for Prefetching with Five Deltas}
\label{alg:kanboost}
\begin{algorithmic}[1]
\STATE \textbf{Input:} Memory access trace $\mathcal{T} = \{a_1, a_2, \dots, a_n\}$
\STATE \textbf{Output:} Prefetched memory addresses $\mathcal{P} = \{p_1, p_2, \dots, p_m\}$

\STATE \textbf{Stage 1: Read Input Data}
\STATE Read memory trace $\mathcal{T}$ and extract access sequence.
\STATE Write predicted prefetch addresses to \texttt{file.address}.

\STATE \textbf{Stage 2: Preprocess Memory Access Trace}
\STATE Split each address $a_i$ into \texttt{page} $p_i$, \texttt{block} $b_i$, and \texttt{offset} $o_i$.
\STATE Compute five deltas for each access:
\[
\Delta_i = (b_i - b_{i-1}, b_{i-1} - b_{i-2}, \cdots , b_{i-4} - b_{i-5})
\]
\STATE Normalize and encode $\{\Delta_i\}$ into a feature set $\mathcal{F}$.

\STATE \textbf{Stage 3: Prepare Training Dataset}
\STATE Define input features $\mathcal{X} = \{(\Delta_i, \Delta_{i-1}, \cdots, \Delta_{i-4})\}$.
\STATE Encode the next delta $\Delta_{i+1}$ as labels $\mathcal{Y}$.
\STATE Split $(\mathcal{X}, \mathcal{Y})$ into training set $(\mathcal{X}_{\text{train}}, \mathcal{Y}_{\text{train}})$ and test set $(\mathcal{X}_{\text{test}}, \mathcal{Y}_{\text{test}})$.

\STATE \textbf{Stage 4: Train \textit{KAN} Model}
\STATE Train a \textit{KAN} model $\mathcal{M}_{\theta}$ with architecture $\mathcal{A}$ and hyperparameters $\theta$.
\STATE Optimize using cross-entropy loss $\mathcal{L}_{\text{test}}$, where:
\[
\mathcal{L} = -\frac{1}{N} \sum_{i=1}^{N} \sum_{c=1}^{C} y_{i,c} \log \hat{y}_{i,c}
\]

\STATE \textbf{Stage 5: Generate Prefetches}
\STATE Predict $\hat{\Delta}_{i+1} = \mathcal{M}_{\theta}(\mathcal{X}_i), \forall i$.
\STATE Compute prefetch addresses: 
\[
p_i = a_i + \hat{\Delta}_{i+1}
\]

\STATE \textbf{Stage 6: Output Prefetches}
\STATE Store prefetched addresses $\mathcal{P}$ in \texttt{file.address}.
\end{algorithmic}
\end{algorithm}

\section{Results and Evaluation methodology}
\subsection{Evaluation setup}
\noindent
We evaluate the proposed KANBoost prefetcher using a fork of Champsim, which is released as part of the ML-DPC Championship 2021 \cite{mlchampsim}.
We implemented this using the Pykan\cite {pykan_github} library, which is a newly created library for implementing KANs. 
System specifications are outlined in 
Table \ref{tab:specifications}.

\begin{table}[htbp]
\centering
\caption{System Specifications Summary with Cache and DRAM Details}
\label{tab:specifications}
\begin{tabular}{|l|l|}
\hline
\textbf{Component}            & \textbf{Specifications} \\ \hline
\textbf{L1 I-Cache}           & 32 KB per core, 8-way, 3-cycle latency \\ \hline
\textbf{L1 D-Cache}           & 48 KB per core, 8-way, 3-cycle latency \\ \hline
\textbf{L2 Cache}             & 512 KB per core, 8-way, 12-cycle latency \\ \hline
\textbf{LLC (Last-Level Cache)} & 12 MB shared, 16-way, 40-cycle latency \\ \hline
\textbf{DRAM}                 & t$_{RP}$=t$_{RCD}$=t$_{CAS}$=22 (DDR4-3200) \\ \hline
\hspace{1em}Channels          & 2 \\ \hline
\hspace{1em}Ranks             & 2 per DIMM \\ \hline
\hspace{1em}Banks             & 16 \\ \hline
\hspace{1em}Rows              & 32K \\ \hline
\hspace{1em}Bandwidth per core & 25.6 GB/s \\ \hline
\end{tabular}
\end{table}

Training on the traces was done offline on a T4 GPU. 
\begin{table}[htbp]
\centering
\caption{System Specifications}
\label{tab:system_specs}
\begin{tabular}{|l|l|}
\hline
\textbf{Component}         & \textbf{Specifications} \\ \hline
\textbf{Processor}         & Apple M1  \\ \hline
\textbf{CPU Cores}        & 8 Cores \\ \hline
\textbf{RAM}               & 8 GB Unified Memory \\ \hline
\textbf{Operating System}  & macOS 13.4.1 \\ \hline
\end{tabular}
\end{table}

\emph{Training configuration}:
The training process for \textbf{KANBoost} was conducted using the \textbf{Adam optimizer}, with \textbf{Cross-Entropy Loss} as the loss function. 
The model's performance was evaluated based on \textbf{training and testing accuracy}. 
Training was carried out for \textbf{1000 steps}. 

Additionally, \textbf{regularization parameters} were applied to improve generalization, with:
\begin{itemize}
    \item Weight regularization parameter: \( \lambda = 0.01 \)
    \item Entropy regularization parameter: \( \lambda_{\text{entropy}} = 8.5 \)
\end{itemize}

\emph{KAN Model architecture}: The KAN model was configured with a layer width of [5, 64, 128], grid size 4, basis functions k=6, random seed 0, and runs on the specified device.
\subsection{Experimental Evaluation}
\noindent
The study evaluates the performance of the KANBoost prefetcher using the \textit{ssp3}, \textit{bc-0}, and \textit{482.sphinx} benchmarks, demonstrating significant improvements in Instructions Per Cycle (IPC), a key measure of processor efficiency. Key findings include:

We see that for the sssp-5 benchmark, KANBoost outperforms the Best offset prefetcher. However for some other benchmarks, KANBoost shows reduction of IPC. This is due to problems in the underlying Machine Learning model, wherein there are restrictions on prefetching sample size, in order to retain hardware feasibility.

We specifically compare our prefetcher with the best offset prefetcher, so as to examine whether KANBoost can be used practically or not.

We compare IPC gains across different prefetchers:
\begin{table}[htbp]
\centering
\caption{Maximum IPC Improvement Comparison}
\label{tab:ipc_comparison}
\begin{tabular}{|l|r|}
\hline
\textbf{Prefetcher} & \textbf{Max IPC Improvement (\%)} \\ \hline
\textbf{KANBoost (Ours)} & 2.5 \\ \hline
\textbf{Best Offset (BO) \cite{Best-offset}} & 60.0 \\ \hline
\textbf{TransforMAP \cite{zhang2021transformap}} & 63.0 \\ \hline
\textbf{Voyager \cite{Voyager}} & 41.6 \\ \hline
\textbf{Drishyam \cite{Drishyam}} & 60.0 \\ \hline
\end{tabular}
\end{table}

\subsection{Viewpoints}
We have observed that the IPC improvement shown by the KAN-based prefetcher is less than the other state-of-the-art prefetchers, such as the best offset prefetcher, \cite{Voyager},\cite{zhang2021transformap} among others.
We attribute this to two reasons:
\begin{itemize}
    \item \textbf{Not considering out-of-page addresses}: To reduce the training time, we considered only future addresses within the same page for this particular work. For the benchmarks used, removing the same-page restrictions poses to be a strong contender for improvement.
    
    \item \textbf{Long-term dependencies}: We used a wide KAN model, which has a bottleneck in capturing memory access patterns. If a recurrent model is employed, it poses to capture the long-term dependencies much better than a wide approach.
\end{itemize}
Conversely, the inference time per sample for the KANBoost prefetcher is about 1000 ns. It is significantly lower than the inference time for the state-of-the art prefetchers, such as \cite{Voyager}. However, works like \cite{Twilight}, which offer even lesser inference time, have formulated the prefetching problem formulation differently, while employing a graph-based heuristic for candidate selection. 
We chose to target the standard prefetching problem formulation so as to ensure a fair comparison, meaning we assume a much larger space for prefetching candidate prediction. 

\begin{table}[htbp]
\centering
\caption{Inference Time Per Sample Comparison}
\label{tab:inference_comparison}
\begin{tabular}{|l|r|}
\hline
\textbf{Prefetcher} & \textbf{Inference Time (ns)} \\ \hline
\textbf{KANBoost (Ours)} & 1000 \\ \hline
Voyager \cite{Voyager} & 18000 \\ \hline
\end{tabular}
\end{table}
This points to an interesting tradeoff while building ML-based prefetchers targeting edge devices: the tradeoff between latency and accuracy, or IPC improvement, and to increase the IPC, one needs to consider long-term dependencies and a wide range of addresses.


\begin{figure*}[htbp]
\centering
\begin{minipage}{0.5\textwidth} 
    \centering
    \includegraphics[width=3 in]{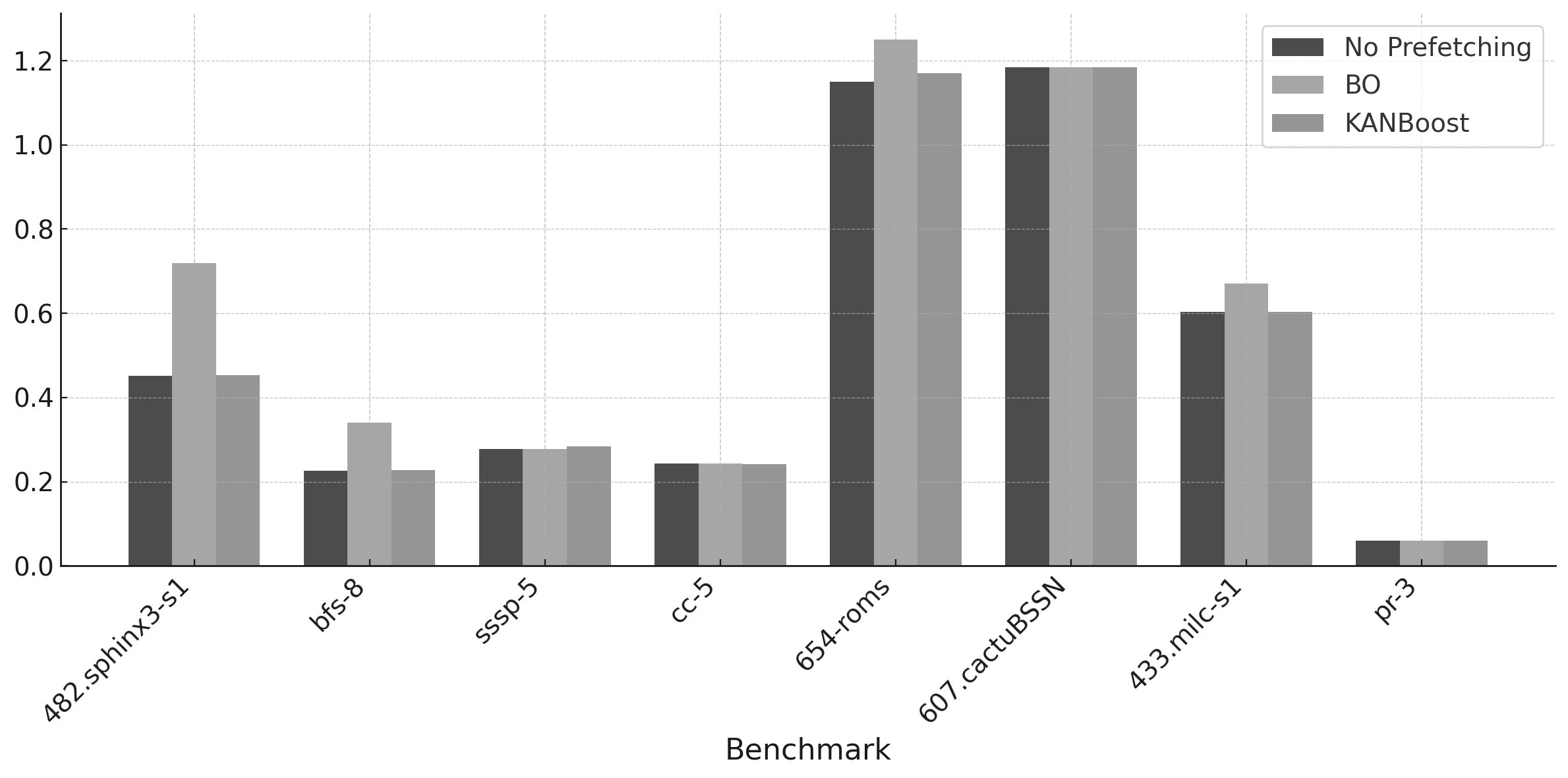} 
    \caption{IPC measures}
    \label{fig:image1}
\end{minipage}%
\hfill
\begin{minipage}{0.5\textwidth} 
    \centering
    \includegraphics[width=3 in]{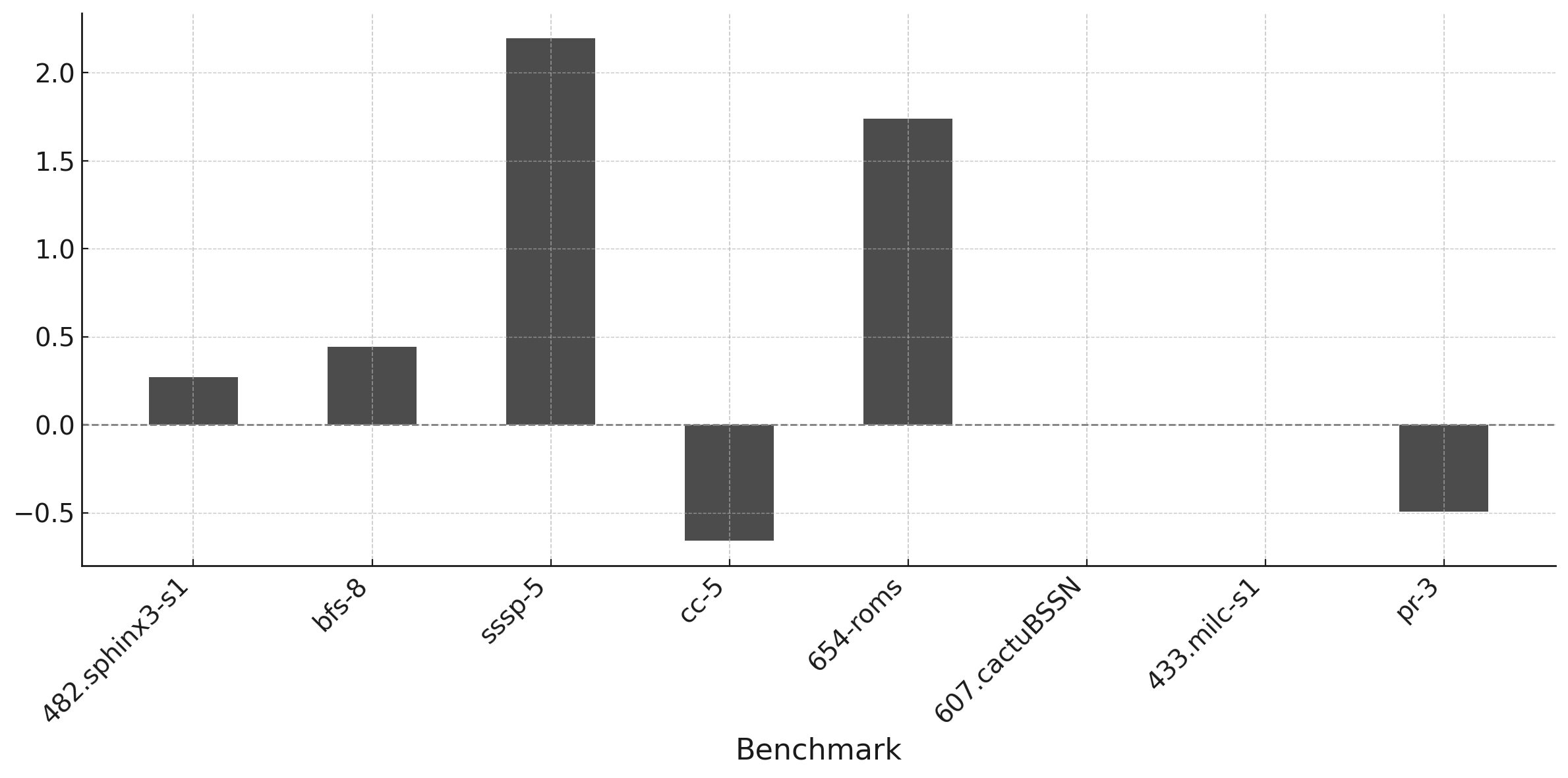} 
    \caption{Normalized IPC improvements compared to a no-prefetching scenario}
    \label{fig:image2}
\end{minipage}
\end{figure*}
\FloatBarrier

\section{Conclusion}
\noindent
We observe that the \textit{KAN}-based prefetcher demonstrates improvement across a variety of benchmarks. Specifically, we achieved  a 2.5\% improvement in Instructions Per Cycle (IPC) compared to a no-prefetching scenario. 

This result points the potential of the Kolmogorov-Arnold representation not only as an effective framework for enabling hardware prefetching but also as a promising approach to evaluate the feasibility of implementing Machine Learning models directly in hardware. Such advancements could lead to substantial gains in memory performance and efficiency, bridging the gap between theoretical computational models and practical hardware implementations.
Future work will focus on achieving greater IPC improvements, surpassing current state-of-the-art ML prefetchers, while targeting lower latency approachable to works like Twilight. Additionally, it is paramount to explore FPGA or ASIC implementations of KANs to ensure actual hardware usage.

\scriptsize

\vspace{-0.8em}
\small
\input{main.bbl}

\end{document}